\documentclass[aps, prd, amsmath, floats, floatfix, twocolumn, nofootinbib, showpacs]{revtex4-1}
\usepackage{graphicx}
\usepackage{color}

\newcommand{\be}{\begin{eqnarray}}
\newcommand{\ee}{\end{eqnarray}}

\begin{document}

\title{Testing the nature of the black hole candidate in GRO~J1655-40\\
with the relativistic precession model}

\author{Cosimo Bambi}
\email{bambi@fudan.edu.cn}

\affiliation{Center for Field Theory and Particle Physics \& Department of Physics, 
Fudan University, 200433 Shanghai, China}

\date{\today}

\begin{abstract}
Quasi-periodic oscillations (QPOs) are a common feature in the X-ray flux of 
stellar-mass black hole candidates, but their exact origin is not yet known. 
Recently, some authors have pointed out that data of GRO~J1655-40 
simultaneously show three QPOs that nicely fit in the relativistic precession 
model. However, they find an estimate of the spin parameter that disagrees
with the measurement of the disk's thermal spectrum. In the present work, I 
explore the possibility of using the relativistic precession model to test the 
nature of the black hole candidate in GRO~J1655-40. If properly understood, 
QPOs may become a quite powerful tool to probe the spacetime geometry 
around black hole candidates, especially if used in combination with other 
techniques. It turns out that the measurements of the relativistic precession 
model and of the disk's thermal spectrum may be consistent if we admit that 
the black hole candidate in GRO~J1655-40 is not of the Kerr type.
\end{abstract}


\maketitle


\section{Introduction}

In 4-dimensional general relativity, uncharged black holes (BHs) are described 
by the Kerr solution and are completely characterized by only two parameters: 
the mass $M$ and the spin angular momentum $J$. This is the result of the 
well-known ``no-hair'' theorem~\cite{hair}. $M$ and $J$ cannot be completely
arbitrary, but they must satisfy the condition for the existence of the event 
horizon $|a| \le M$, where $a = J/M$ is the spin parameter. Astrophysical BHs,
if they exist, should be well described by the Kerr metric: initial deviations from 
the Kerr geometry are expected to be quickly radiated away through the emission 
of gravitational waves~\cite{price}, an initially non-vanishing electric charge 
would be shortly neutralized in their highly ionized environment~\cite{bdp}, while 
the presence of the accretion disk is completely negligible in most cases.

Astronomical observations have discovered at least two classes of BH candidates:
stellar-mass objects in X-ray binary systems with a mass $M \approx 5 - 20$~$M_\odot$, 
and super-massive bodies in galactic nuclei with a mass $M \sim 10^5 - 
10^9$~$M_\odot$~\cite{nara}. All these objects are thought to be the Kerr BHs
of general relativity, but their actual nature is still to be verified. Robust measurements 
of the masses of these objects can be obtained from dynamical methods, 
by studying the orbital motion of gas or of individual stars around them. Such a
measurements are the main argument to support the Kerr BH hypothesis, 
because these objects are so heavy that they cannot be explained otherwise 
without introducing new physics. The non-observation of electromagnetic radiation 
emitted by the possible surface of these objects may also be interpreted as an 
indication for the existence of an event horizon~\cite{horizon1} (but see~\cite{horizon2}). 
However, there is no evidence that the spacetime geometry around them is 
described by the Kerr solution.

The nature of astrophysical BH candidates may be potentially tested with the already 
available X-ray data, because the features of the electromagnetic radiation emitted 
by the gas of the accretion disk can provide information on the spacetime geometry
around these compact objects (for a review, see e.g.~\cite{review}). The study of the 
disk's thermal spectrum (continuum-fitting method)~\cite{cfm} and the analysis of 
the profile of the broad K$\alpha$ iron line~\cite{ka} are today the only two relatively 
mature techniques to probe the metric around BH candidates. They have been
developed to infer the spin parameter of these objects under the assumption of the
Kerr spacetime, but more recently they have been extended to check the Kerr
background~\cite{cb1,cb2,jp,torres}. The main problem to test the Kerr BH 
paradigm with these techniques is that it is extremely difficult to get independent 
estimates of the spin parameter and of possible deviations from the Kerr solution. 
In other words, one can usually only constrain a combination of the spin and of 
possible deviations, because the properties of the radiation emitted by the gas 
in the accretion disk around a non-Kerr object with a certain spin can be very similar 
to the ones produced in the spacetime of a Kerr BH with different spin. The possibility 
of combining the continuum-fitting method and the iron line analysis for the same 
object has been discussed in Ref.~\cite{cb3}. For some BH solutions, the 
combination of the two approaches is not very helpful and the Kerr metric cannot
be unambiguously tested. In other BH backgrounds, the study of the disk's thermal
spectrum and the analysis of the iron line profile of a specific source can do the
job, but quite accurate measurements are usually necessary. 
The possibility of using the estimate of the power of transient or 
steady jets with the measurements from the continuum-fitting method has been 
explored in Ref.~\cite{jet}. While the approach seems to be promising, the 
mechanisms responsible for the formation of these jets are not known and different 
interpretations lead to different conclusions. In the future, high resolution sub-mm 
observations will be hopefully able to detect the ``shadow'' of nearby super-massive 
BH candidates, opening a new window to test the spacetime geometry around 
these objects~\cite{shadow}.

Quasi-periodic oscillations (QPOs) are a very promising tool to get precise
information on the spacetime geometry around stellar-mass BH candidates.
They are seen as peaks in the X-ray power density spectra 
of the source. At present, however, the exact physical mechanism responsible 
for the production of these QPOs is not understood and several different scenarios 
have been proposed, including relativistic precession models~\cite{rpm},
diskoseismology models~\cite{dm}, resonance models~\cite{rm}, and 
$p$-mode oscillations of an accretion torus~\cite{rezz}. In most scenarios, 
the frequencies of the QPOs are directly related to the characteristic orbital 
frequencies of a test-particle, which are determined only by the background 
metric and are independent of the complicated astrophysical processes of 
the accretion. While such a correlation with the fundamental frequencies
of the spacetime may sound quite artificial at first, it is possible to show that there
is indeed a direct relation between these frequencies and the ones of the oscillation 
modes of the fluid accretion flow. The significant advantage of the use of QPOs
with respect to other techniques is that the frequencies of the QPOs can be 
measured with high accuracy, and therefore they can potentially be used to
get very precise measurements of the parameters of the spacetime geometry
of the compact object. Attempts to use the QPOs to test the Kerr metric around
BH candidates are reported in~\cite{qpo}. However, since we do not know the
exact mechanism responsible for these oscillations, such a powerful approach
cannot yet be used. Different models relate the fundamental frequencies of the 
background and the observed frequencies of the QPOs in a different way,
and current X-ray data are not able to select the correct model and rule out the
others.

Very recently, some authors have pointed out that the X-ray data of GRO~J1655-40
nicely fit in the relativistic precession model~\cite{belloni}. The key-point is that this
source is the only one for which three simultaneous QPOs have been observed.
In the Kerr spacetime, the three fundamental frequencies of the background metric
(orbital frequency, radial epicyclic frequency, and vertical epicyclic frequency)
depend on the radius $r$, the BH mass $M$, and the BH spin parameter $a$.
Assuming that the three observed QPOs are generated at the same radius $r$,
one has a system of three equations with three unknown variables ($r$, $M$, and $a$).
The system of the equations can therefore be solved to find $r$, $M$, and $a$,
which can be determined with a quite small uncertainty due to the high precision
of the measurement of the frequencies. The authors of Ref.~\cite{belloni} find
that the inferred value of $M$ is in agreement with the value obtained by dynamical
methods in Ref.~\cite{beer}. In support of the relativistic precession model, 
the authors of Ref.~\cite{belloni} show also that the X-ray data of GRO~J1655-40
with two simultaneous QPOs can be nicely interpreted as two of the three 
frequencies generated at radii $r$ larger than the one found in the data with three
frequencies. However, their spin measurement is not consistent with the one
obtained from the continuum-fitting method in Ref.~\cite{shafee}.

The aim of the present paper is to investigate the possibility of using the data and 
the interpretation of Ref.~\cite{belloni} to test the spacetime geometry around 
the BH candidate in GRO~J1655-40. For this purpose, it is convenient to consider 
a metric more general than the Kerr one, with one (or more) deformation parameter(s). 
The latter is used to measure possible deviations from the Kerr background, which 
must be recovered when the deformation parameter vanishes. Now one needs an
independent measurement of the mass of the BH candidate, so that it is possible
to solve the system of equations of the
three frequencies to find the three unknown quantities ($r$, $a$, and the deformation 
parameter). The result is an allowed region on the spin-deformation parameter plane,
just like the author of Ref.~\cite{belloni} find an allowed region on the mass-spin plane.
The strong correlation between the spin and possible deviations from the Kerr
solution found with other approaches is present even here, but the size of the allowed
region is much smaller, supporting the idea that, if properly understood, QPOs can 
be a very powerful tool to probe the spacetime geometry around BH candidates.  
In order to check the validity of this result, the latter is compared with the allowed 
region on the spin-deformation parameter plane inferred from the study of the
disk's thermal spectrum of GRO~J1655-40~\cite{shafee}. It turns out that the
disagreement between the two measurements found in the Kerr metric cannot be 
solved if we believe in the mass measurement of Ref.~\cite{beer}. However, a
different measurement of the mass of the BH candidate in GRO~J1655-40 is
reported in~\cite{shafee}. If we believe in the mass measurement of this work, the 
one found in the Kerr background with the relativistic precession model 
in~\cite{belloni} is wrong, while it is possible to reconcile the QPO measurement
of the spin with the disk's thermal spectrum analysis if we allow for deviations 
from the Kerr geometry. In the latter case, the non-vanishing deformation parameter
would be compatible with the one inferred in the second paper in~\cite{jet} from the 
combination of
the measurements of the disk spectrum and the estimates of the power of steady 
jets. While that may be accidental, if the relativistic precession model turns out
to be right we may suspect that the continuum-fitting method regularly overestimates
the spin parameter or even speculate on the violation of the Kerr BH paradigm.

The content of the paper is as follows. In Section~\ref{s-rpm}, I briefly review
the relativistic precession model and the results of Ref.~\cite{belloni}, valid in the
Kerr background. In Section~\ref{s-test}, I apply this approach to the rotating
Bardeen BH metric~\cite{m-b} and to the Johannsen-Psaltis background~\cite{m-jp}
to find an allowed region on the spin-deformation parameter plane. In
Section~\ref{s-d}, I discuss these results, which are also compared with the
constraints that can be obtained from the continuum-fitting method. Summary 
and conclusions are reported in Section~\ref{s-c}. Throughout the paper, I use 
units in which $G_{\rm N} = c = 1$, unless stated otherwise.

\begin{figure*}
\begin{center}
\hspace{-0.5cm}
\includegraphics[type=pdf,ext=.pdf,read=.pdf,width=8.5cm]{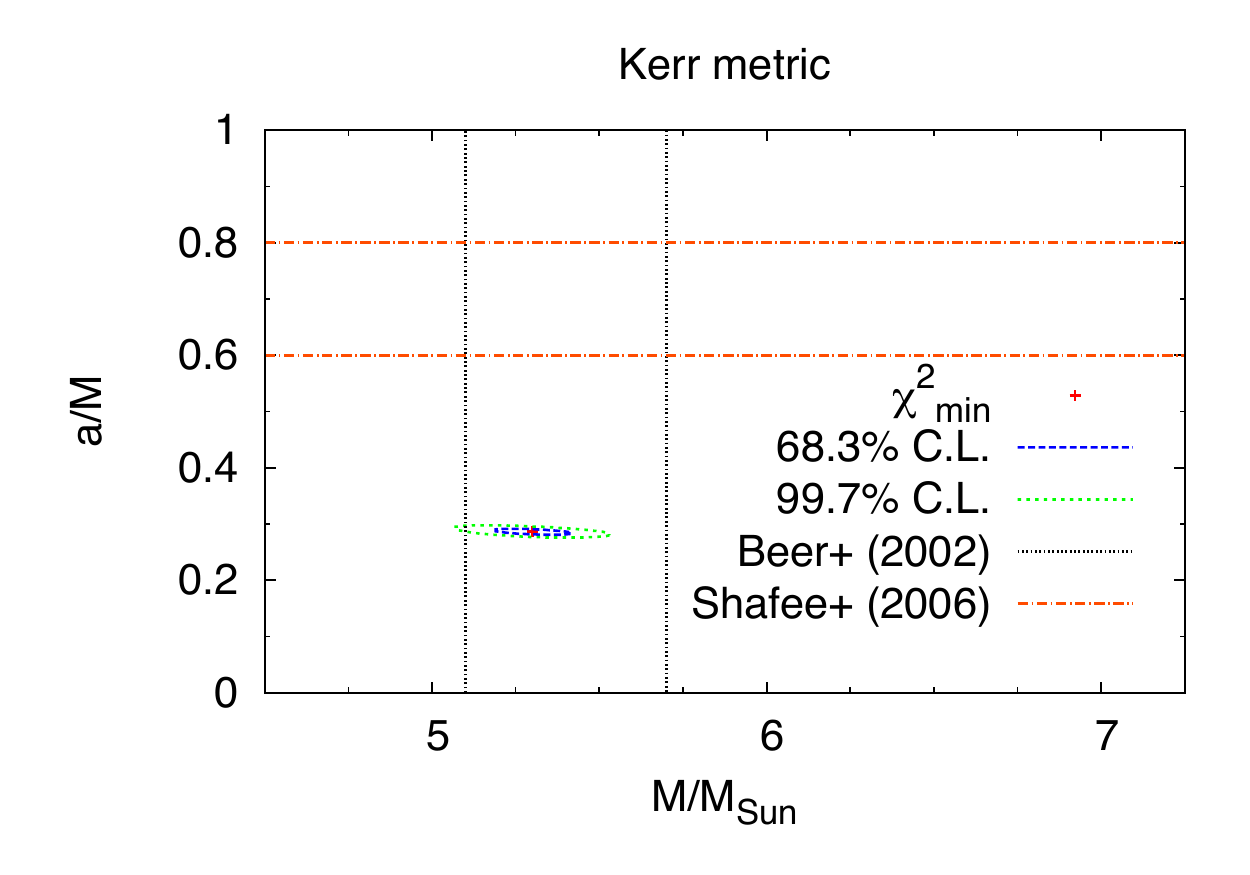}
\includegraphics[type=pdf,ext=.pdf,read=.pdf,width=8.5cm]{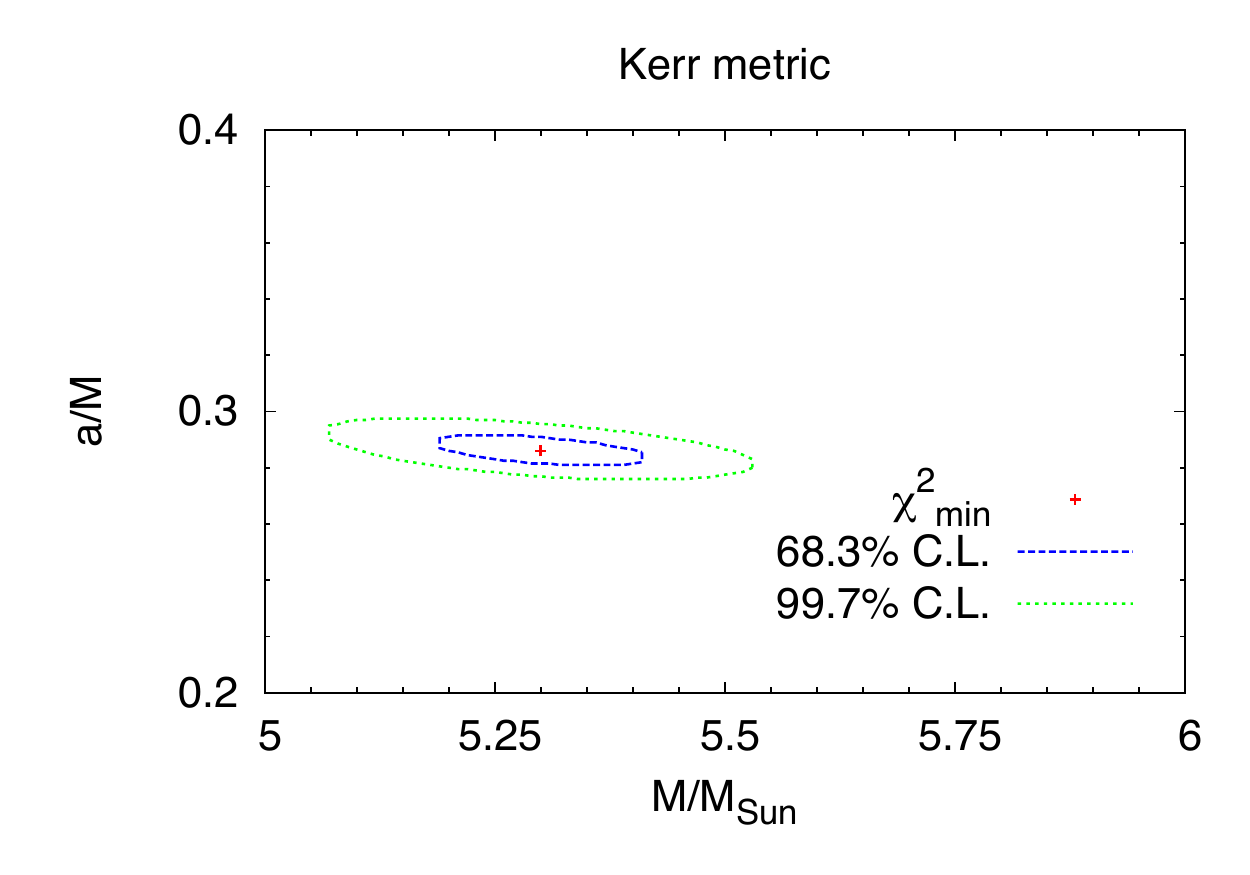}
\end{center}
\vspace{-0.5cm}
\caption{Estimate of the mass $M$ and of the spin parameter $a/M$ of the
BH candidate in GRO~J1655-40 with the relativistic precession model of
Ref.~\cite{belloni} and under the assumption of the Kerr background. With the
approach discussed in Section~\ref{s-rpm}, the result is $M/M_\odot = 5.30 \pm 0.11$ 
and $a/M = 0.286 \pm 0.006$ (68.3\% C.L.). The vertical black thin-dotted lines
are the boundaries of the optical measurement of the mass of this object found in 
Ref.~\cite{beer}, while the orange dashed-dotted curve is the boundary of
the allowed region for the spin parameter via the continuum-fitting method 
obtained in Ref.~\cite{shafee}. The right panel is just the enlargement of the left 
one. See the text for more details.}
\label{fig1}
\end{figure*}

\section{The relativistic precession model \label{s-rpm}}

The relativistic precession model was originally proposed to explain QPOs
in low-mass X-ray binaries with a neutron star and was then extended to
systems with stellar-mass BH candidates~\cite{rpm}. It does not really
explain the origin of the QPOs, but it simply relates the observed frequencies
of the QPOs with the three fundamental frequencies of the background metric.
The latter are the Keplerian frequency of equatorial circular orbits (orbital 
frequency $\nu_\phi$) and the frequencies of small perturbations along the radial 
and vertical direction around the equatorial circular orbit (respectively the
radial epicyclic frequency $\nu_r$ and the vertical epicyclic frequency 
$\nu_\theta$). In the Kerr metric, these frequencies can be written in
analytic form and are given by
\be
\nu_\phi &=& \left(\frac{1}{2\pi}\right)
\frac{M^{1/2}}{r^{3/2} \pm a M^{1/2}} \, , \\
\nu_r &=& \nu_\phi
\left( 1 - \frac{6 M}{r} \pm \frac{8 a M^{1/2}}{r^{3/2}}
- \frac{3 a^2}{r^2} \right) \, , \\
\nu_\theta &=& \nu_\phi
\left( 1 \mp \frac{4 a M^{1/2}}{r^{3/2}} + \frac{3 a^2}{r^2} \right) \, ,
\ee
where the upper (lower) sign is for the case of corotating (counterrotating)
orbits. From these three frequencies, one can find the periastron precession 
frequency $\nu_{\rm p}$ and the nodal precession frequency $\nu_{\rm n}$,
given by
\be
\nu_{\rm p} &=& \nu_\phi - \nu_r \, , \nonumber\\
\nu_{\rm n} &=& \nu_\phi - \nu_\theta \, .
\ee
All these frequencies depend on three parameters; that is, the radius of
the orbit $r$ and the two parameters of the background geometry, the
BH mass $M$ and the BH spin parameter $a$.

In X-ray binaries with a BH candidate, observations have detected low-frequency
QPOs of different nature (type-A, type-B, and type-C) in the range $\sim 0.1-30$~Hz,
and high-frequency QPOs at $\sim 100-400$~Hz. The latter may be seen in pairs
and they are therefore called lower and upper high-frequency QPOs.
The crucial point is to find the correct relation between the fundamental 
frequencies of the background metric and the ones of the observed QPOs. In 
Ref.~\cite{belloni}, the authors propose the following interpretation (which is not 
exactly the original proposal of the relativistic precession model 
in~\cite{rpm}). The low-frequency type-C QPO $\nu_{\rm C}$ would correspond
to the nodal precession frequency $\nu_{\rm n}$, while the lower high-frequency 
QPO $\nu_{\rm L}$ and the upper high-frequency QPO $\nu_{\rm U}$ would be
associated, respectively, to the periastron precession frequency $\nu_{\rm p}$
and to the orbital frequency $\nu_\phi$:
\be
\nu_{\rm C} = \nu_{\rm n} \, , \quad
\nu_{\rm L} = \nu_{\rm p} \, , \quad
\nu_{\rm U} = \nu_\phi \, .
\ee

The case of the BH candidate in GRO~J1655-40 is special, because it is the
only BH system for which we have data with three simultaneous QPOs. 
The low-frequency type-C QPO used in~\cite{belloni} was identified in
Ref.~\cite{motta}, while the two high-frequency QPOs were found in~\cite{strohmayer}.
Since one sees simultaneously the three frequencies, it is possible to argue that
they may be associated to oscillations of the fluid flow at the same radial 
coordinate. In this way, one can solve the system of equations of the three
frequencies ($\nu_{\rm C}$, $\nu_{\rm L}$, and $\nu_{\rm U}$) to find the
three unknown variables ($r$, $M$, and $a$). The system of equations
cannot be solved analytically and therefore one has to find the three
parameters numerically. Here I use a different approach with respect to
Ref.~\cite{belloni} and I compute the $\chi$-square as follows:
\be\label{eq-chi20}
\chi^2_0 (r,M,a) &=& 
\frac{\left( \nu_{\rm C} - \nu_{\rm n} \right)^2}{\sigma^2_{C}}
+ \frac{\left( \nu_{\rm L} - \nu_{\rm p} \right)^2}{\sigma^2_{L}}
+ \nonumber\\ &&
+ \frac{\left( \nu_{\rm U} - \nu_\phi \right)^2}{\sigma^2_{U}} \, .
\ee
For GRO~J1655-40, we have~\cite{belloni}
\be
\begin{matrix}
\nu_{\rm C} = 17.3 \; {\rm Hz} \, , && \sigma_{C} = 0.1 \; {\rm Hz} \, , \\
\nu_{\rm L} = 298 \; {\rm Hz} \, , && \sigma_{L} = 4 \; {\rm Hz} \, , \\
\nu_{\rm U} = 441 \; {\rm Hz} \, , && \sigma_{U} = 2 \; {\rm Hz} \, . \\
\end{matrix}
\ee
The minimum of $\chi^2_0$ (which should be zero in this case, as the
system of equations has always a solution) gives the estimate of $r$, $M$,
and $a$, while the intervals defined by 
$\chi^2_0 = \chi^2_{0 , \, {\rm min}} + \Delta \chi^2_0$ give the ranges of
$r$, $M$, and $a$ at the confidence level (C.L.) set by $\Delta \chi^2_0$.
In the case of three degrees of freedom, $\Delta \chi^2_0 = 3.53$, 8.03,
and 14.16 correspond, respectively, to 68.3\%, 95.4\%, and 99.7\% C.L.,
which are the probability intervals designated as 1, 2, and 3 standard 
deviation limits.

Following this procedure, one finds the plots in Fig.~\ref{fig1}. The final 
result for the mass and the spin parameter is $M/M_\odot = 5.30 \pm 0.11$ 
and $a/M = 0.286 \pm 0.006$ (68.3\% C.L.). The estimate of the mass is
consistent with the value inferred by optical observations in Ref.~\cite{beer},
$M/M_\odot = 5.4 \pm 0.3$,
which corresponds to the black thin-dotted lines in the left panel in Fig.~\ref{fig1}. 
However, in the literature there is also another mass measurement of the
BH candidate in GRO~J1655-40, $M/M_\odot = 6.3 \pm 0.3$, reported 
in~\cite{shafee}.
The orange dashed-dotted curve corresponds instead to the measurement
of the spin parameter inferred via the continuum-fitting method in Ref.~\cite{shafee},
$a/M = 0.7 \pm 0.1$\footnote{Actually, the measurement in Ref.~\cite{shafee} is 
$a/M = 0.70 \pm 0.05$ at 1-sigma, but since it was one of the first measurements
obtained with the continuum-fitting method by the CfA group, in later studies using
this result the uncertainty has been conservatively doubled by the same authors.}.
Such a measurement does depend on the BH mass $M$, but in Fig.~\ref{fig1}
I show only the best estimate for $a/M$ assuming that the mass (which is an 
input parameter in the continuum-fitting method) obtained by optical measurements
is correct. 
In Ref.~\cite{shafee}, the authors use $M/M_\odot = 6.3 \pm 0.3$,
not the one of Ref.~\cite{beer}, but the effect on the estimate of the spin is not large
and cannot solve the disagreement between the relativistic precession model and
the continuum-fitting method.
The measurement of the frequencies of QPOs can
potentially provide very precise estimates of the mass $M$ and the spin parameter 
$a$ with respect to other techniques. However, the measurement
from the relativistic precession model and the continuum-fitting method provide
inconsistent results, which means that either one of the two approaches 
provides an erroneous value of the spin parameter $a/M$, or both. In the next
sections, I will check if the two techniques can give consistent results if we allow 
for deviations from the Kerr background.

\section{Testing the Kerr nature of GRO~J1655-40 \label{s-test}}

A common approach to test the nature of astrophysical BH candidates and
constrain possible deviations from the Kerr solution is to consider a more
general background, which includes the Kerr metric as a special case. In addition
to the mass $M$ and the spin parameter $a$, the spacetime geometry is
characterized by at least one more parameter, which is used to measure
possible deviations from the Kerr background. The idea is to infer $M$, $a$
and such a deformation parameter from observational data and check if
the latter require a vanishing deformation parameter; that is, the compact
object is a Kerr BH. On the contrary, if it turns out that observations require
a non-vanishing deformation parameter, the BH candidate may not be 
a Kerr BH.

Let us now revise the relativistic precession model in a generic stationary, 
axisymmetric, and asymptotically flat spacetime. The line element of the 
spacetime can be written in the canonical form
\be
ds^2 &=& g_{tt} dt^2 + g_{rr}dr^2 + g_{\theta\theta} d\theta^2 
+ 2g_{t\phi}dt d\phi + \nonumber\\ && + g_{\phi\phi}d\phi^2 \, ,
\ee
where the metric components are independent of the $t$ and $\phi$ coordinates, 
which implies the existence of two constants of motion: the conserved specific 
energy at infinity, $E$, and the conserved $z$-component of the specific angular 
momentum at infinity, $L_z$. This fact allows to write the $t$- and 
$\phi$-component of the 4-velocity of a test-particle as 
\be
\dot{t} = \frac{E g_{\phi\phi} + L_z g_{t\phi}}{
g_{t\phi}^2 - g_{tt} g_{\phi\phi}} \, , \quad 
\dot{\phi} = - \frac{E g_{t\phi} + L_z g_{tt}}{
g_{t\phi}^2 - g_{tt} g_{\phi\phi}} \, .
\ee
From the conservation of the rest-mass, $g_{\mu\nu}\dot{x}^\mu \dot{x}^\nu = -1$,
we can write
\be
g_{rr}\dot{r}^2 + g_{\theta\theta}\dot{\theta}^2
= V_{\rm eff}(r,\theta,E,L_z) \, ,
\ee
where the effective potential $V_{\rm eff}$ is given by
\be
V_{\rm eff} = \frac{E^2 g_{\phi\phi} + 2 E L_z g_{t\phi} + L^2_z 
g_{tt}}{g_{t\phi}^2 - g_{tt} g_{\phi\phi}} - 1  \, .
\ee
Circular orbits on the equatorial plane are located at the zeros and the turning 
points of the effective potential: $\dot{r} = \dot{\theta} = 0$, which implies 
$V_{\rm eff} = 0$, and $\ddot{r} = \ddot{\theta} = 0$, requiring respectively 
$\partial_r V_{\rm eff} = 0$ and $\partial_\theta V_{\rm eff} = 0$. From these 
conditions, one can obtain the orbital angular velocity $\Omega_\phi = d\phi/dt$, 
$E$, and $L_z$ of the test-particle:
\be
\Omega_\phi &=& \frac{- \partial_r g_{t\phi} 
\pm \sqrt{\left(\partial_r g_{t\phi}\right)^2 
- \left(\partial_r g_{tt}\right) \left(\partial_r 
g_{\phi\phi}\right)}}{\partial_r g_{\phi\phi}} \, , \\
E &=& - \frac{g_{tt} + g_{t\phi}\Omega_\phi}{
\sqrt{-g_{tt} - 2g_{t\phi}\Omega_\phi - g_{\phi\phi}\Omega^2_\phi}} \, , \\
L_z &=& \frac{g_{t\phi} + g_{\phi\phi}\Omega_\phi}{
\sqrt{-g_{tt} - 2g_{t\phi}\Omega_\phi - g_{\phi\phi}\Omega^2_\phi}} \, ,
\ee
where in $\Omega_\phi$ the sign is $+$ ($-$) for corotating (counterrotating) orbits. 
The orbital frequency is simply $\nu_\phi = \Omega_\phi/2\pi$. The orbits are 
stable under small perturbations if $\partial_r^2 V_{\rm eff} \le 0$ and 
$\partial_\theta^2 V_{\rm eff} \le 0$.

The radial and vertical epicyclic frequencies can be quickly computed by considering
small perturbations around circular equatorial orbits, respectively along the radial and
vertical direction. If $\delta_r$ and $\delta_\theta$ are the small displacements around
the mean orbit (i.e. $r = r_0 + \delta_r$ and $\theta = \pi/2 + \delta_\theta$), 
we find they are governed by the following differential equations
\be\label{eq-o1}
\frac{d^2 \delta_r}{dt^2} + \Omega_r^2 \delta_r &=& 0 \, , \\
\frac{d^2 \delta_\theta}{dt^2} + \Omega_\theta^2 \delta_\theta &=& 0 \, ,
\label{eq-o2}
\ee
where
\be\label{eq-or}
\Omega^2_r &=& - \frac{1}{2 g_{rr} \dot{t}^2} 
\frac{\partial^2 V_{\rm eff}}{\partial r^2} \, , \\
\Omega^2_\theta &=& - \frac{1}{2 g_{\theta\theta} \dot{t}^2} 
\frac{\partial^2 V_{\rm eff}}{\partial \theta^2} \, .
\label{eq-ot}
\ee
The radial epicyclic frequency is thus $\nu_r = \Omega_r/2\pi$ and the vertical one 
is $\nu_\theta = \Omega_\theta/2\pi$.

\begin{figure*}
\begin{center}
\hspace{-0.5cm}
\includegraphics[type=pdf,ext=.pdf,read=.pdf,width=8.5cm]{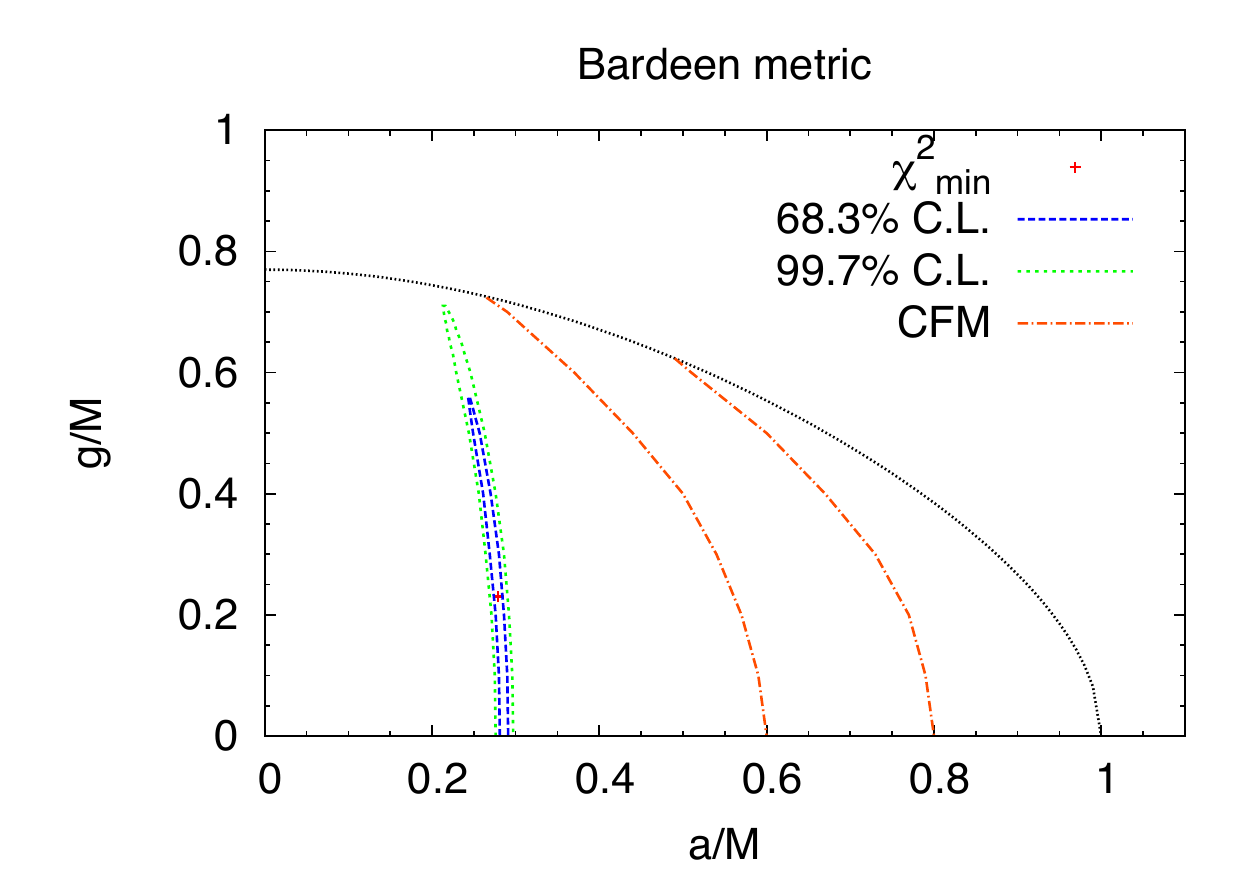}
\includegraphics[type=pdf,ext=.pdf,read=.pdf,width=8.5cm]{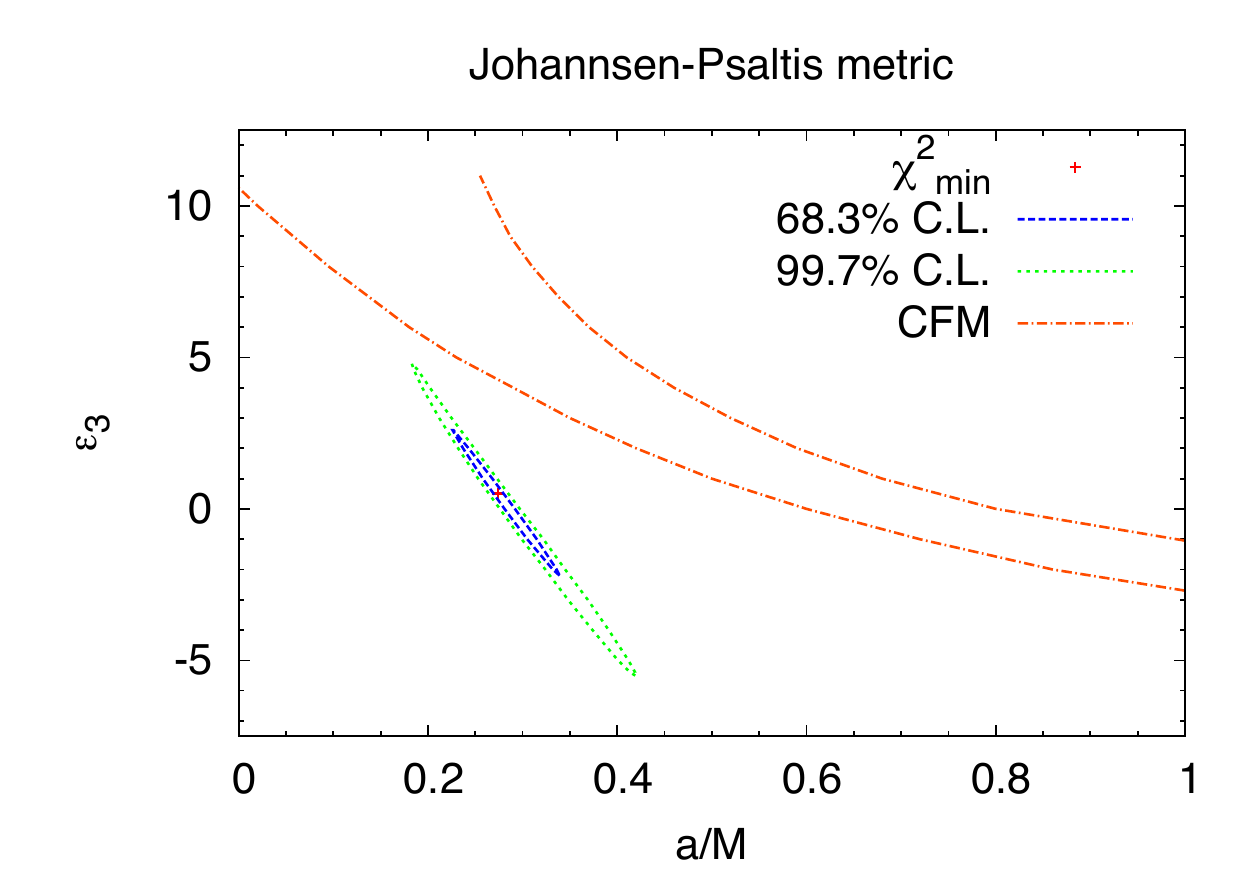}
\end{center}
\vspace{-0.5cm}
\caption{Constraints on the spacetime geometry around the BH candidate in 
GRO~J1655-40 with the relativistic precession model (blue dashed and green 
dotted lines) and the continuum-fitting method (orange dashed-dotted lines).
The relativistic precession model assumes the mass measurement 
$M/M_\odot = 5.4 \pm 0.3$ reported in Ref.~\cite{beer}.
Left panel: Bardeen background, where the black thin-dotted line is the 
boundary separating BHs from horizonless objects. Right panel: Johannsen-Psaltis 
background with deformation parameter $\epsilon_3$. See the text for more details.}
\label{fig2}
\end{figure*}

As first example of non-Kerr background, we can consider the Bardeen BH 
metric~\cite{m-b}. In Boyer-Lindquist coordinates, the non-vanishing metric 
coefficients are
\be\label{eq-metric}
&&g_{tt} = - \left(1 - \frac{2 m r}{\Sigma}\right) \, , \quad
g_{t\phi} = - \frac{2 a m r \sin^2 \theta}{\Sigma} \, , \nonumber\\
&&g_{\phi\phi} = \left(r^2 + a^2 + \frac{2 a^2 m r \sin^2\theta}{\Sigma}\right)
\sin^2\theta \, , \nonumber\\
&&g_{rr} = \frac{\Sigma}{\Delta}\, ,
\quad g_{\theta\theta} = \Sigma \, ,
\ee
where
\be
&&\Sigma = r^2 + a^2 \cos^2\theta \, , \quad
\Delta = r^2 - 2 m r + a^2 \, , \nonumber\\
&&m = M \left(\frac{r^2}{r^2 + g^2}\right)^{3/2} \, .
\ee
$g$ can be interpreted as the magnetic charge of a non-linear electromagnetic
field or just as a quantity introducing a deviation from the Kerr metric. The position 
of the even horizon is given by the larger root of $\Delta = 0$ and therefore there 
is a bound on the maximum value of the spin parameter, above which there are 
no BHs. The maximum value of $a$ is $M$ for $g/M = 0$ (Kerr case), and decreases 
as $g/M$ increases. The black thin-dotted curve in the left panel of Fig.~\ref{fig2}
is the boundary separating BH solutions (left bottom corner) and horizonless 
solutions (right top corner) on the plane $(a/M, g/M)$. Since the horizonless
solutions are likely very unstable objects with a short lifetime due to the 
ergoregion instability, they can be safely ignored.

Now we have three equations for $\nu_{\rm n}$, $\nu_{\rm p}$, and $\nu_\phi$ 
and four unknown variables ($r$, $M$, $a$, and $g$). In order to solve the
system, we need an independent estimate of the mass $M$. In this case, the 
$\chi$-square becomes
\be\label{eq-chi2}
\chi^2 (r,a,g) = \min_M \left[ \chi^2_0 + 
\frac{\left( M - M_{\rm opt} \right)^2}{\sigma^2_M} \right] \, ,
\ee
where $\chi_0^2$ is given in Eq.~(\ref{eq-chi20}) and the three degrees
of freedom are now $r$, $a$, and $g$. For $M_{\rm opt} = 5.4$~$M_\odot$ 
and $\sigma_M = 0.3$~$M_\odot$~\cite{beer}, the result is shown in
the left panel of Fig.~\ref{fig2}, where the constraints on the spin and
on possible deviations from the Kerr solutions are
\be
a/M &=& 0.279^{+0.012}_{-0.036} \, , \nonumber\\
g/M &<& 0.56 \, ,
\ee
at the 68.3\% C.L.

To check the genericity of this result found in the specific case of the Bardeen 
BH solution, it is convenient to repeat the same exercise with a different 
background metric. As second example, now I consider the Johannsen-Psaltis 
metric, whose non-vanishing metric coefficients in Boyer-Lindquist coordinates 
are~\cite{m-jp}:
\be\label{eq-jp}
g_{tt} &=& - \left(1 - \frac{2 M r}{\Sigma}\right) (1 + h) \, , \nonumber\\
g_{t\phi} &=& - \frac{4 a M r \sin^2\theta}{\Sigma} (1 + h) \, , \nonumber\\
g_{\phi\phi} &=& \sin^2\theta \left(r^2 + a^2 
+ \frac{2 a^2 M r \sin^2\theta}{\Sigma} \right) + \nonumber\\&&
+ \frac{a^2 (\Sigma + 2 M r) \sin^4\theta}{\Sigma} h \, , \nonumber\\
g_{rr} &=& \frac{\Sigma (1 + h)}{\Delta + a^2 h \sin^2\theta } \, ,
\qquad g_{\theta\theta} = \Sigma \, ,
\ee
where
\be
&&\Sigma = r^2 + a^2 \cos^2\theta \, , \quad
\Delta = r^2 - 2 M r + a^2 \, , \nonumber\\
&&h = \sum_{k = 0}^{\infty} \left(\epsilon_{2k}
+ \frac{M r}{\Sigma} \epsilon_{2k+1} \right)
\left(\frac{M^2}{\Sigma}\right)^k \, .
\ee
Such a metric has an infinite number of deformation parameters $\epsilon_k$
($k = 0$, 1, 2, ...). However, $\epsilon_0 = \epsilon_1 = 0$ in order to recover the
correct Newtonian limit, while $\epsilon_2$ is strongly constrained by Solar 
System experiments~\cite{m-jp}. For the sake of simplicity, I will consider
the case of a single deformation parameter $\epsilon_3$ and set to zero all the
others. One can then define the counterpart of the $\chi$-square in Eq.~(\ref{eq-chi2})
\be
\chi^2 (r,a,\epsilon_3) = \min_M \left[ \chi^2_0 + 
\frac{\left( M - M_{\rm opt} \right)^2}{\sigma^2_M} \right] \, .
\ee
With the mass measurement of Ref.~\cite{beer}, the 
result is the plot in the right panel of Fig.~\ref{fig2}. There is a quite
pronounced correlation between the estimate of the spin and the deformation
parameters, as shown by the thin but quite inclined position of the allowed 
region. The constraints are
\be
a/M &=& 0.27^{+0.06}_{-0.05} \, , \nonumber\\
\epsilon_3 &=& 0.5^{+2.1}_{-2.7} \, ,
\ee
at the 68.3\% C.L.

\begin{figure*}
\begin{center}
\hspace{-0.5cm}
\includegraphics[type=pdf,ext=.pdf,read=.pdf,width=8.5cm]{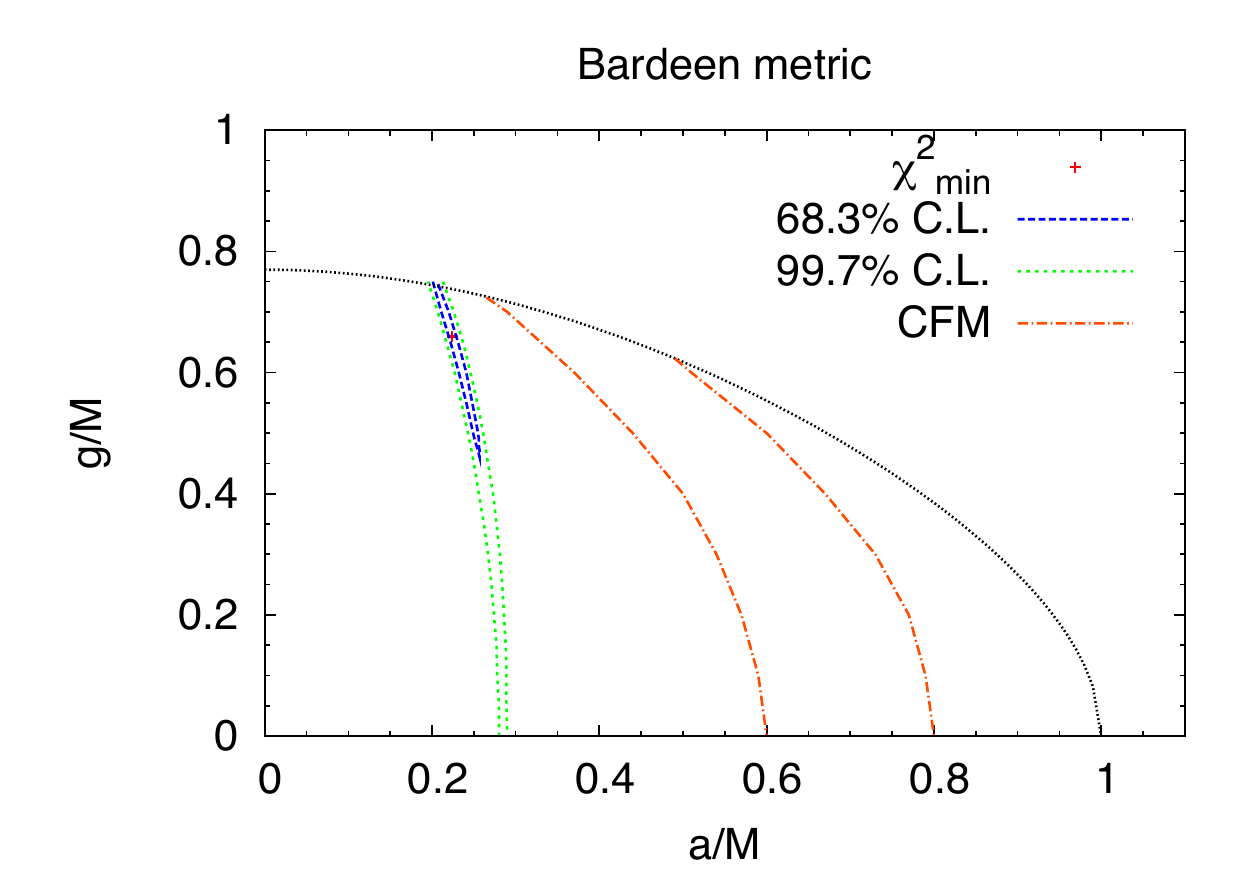}
\includegraphics[type=pdf,ext=.pdf,read=.pdf,width=8.5cm]{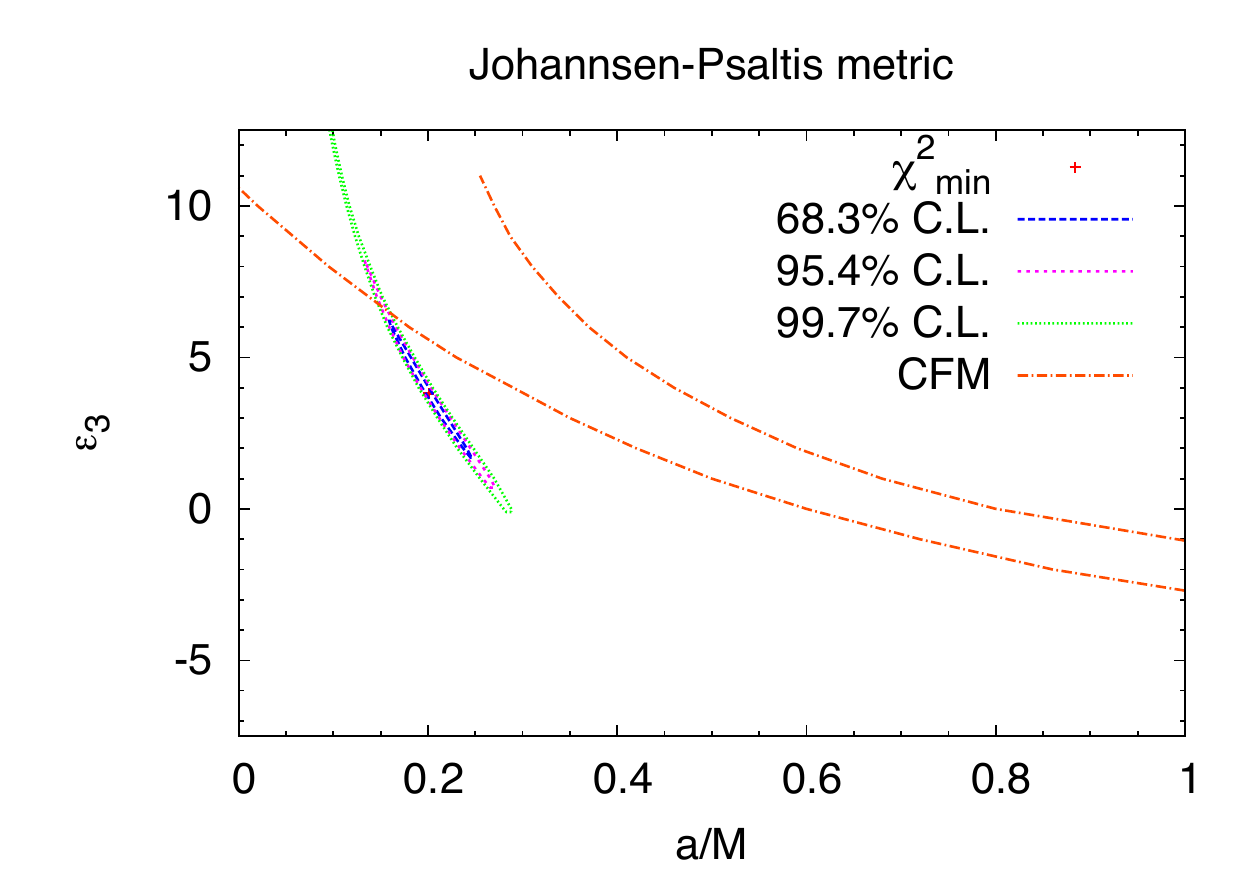}
\end{center}
\vspace{-0.5cm}
\caption{As in Fig.~\ref{fig2}, in the case in which the relativistic precession model 
uses the mass measurement $M/M_\odot = 6.3 \pm 0.3$ reported in Ref.~\cite{shafee}.}
\label{fig2bis}
\end{figure*}

\begin{figure*}
\begin{center}
\hspace{-0.5cm}
\includegraphics[type=pdf,ext=.pdf,read=.pdf,width=8.5cm]{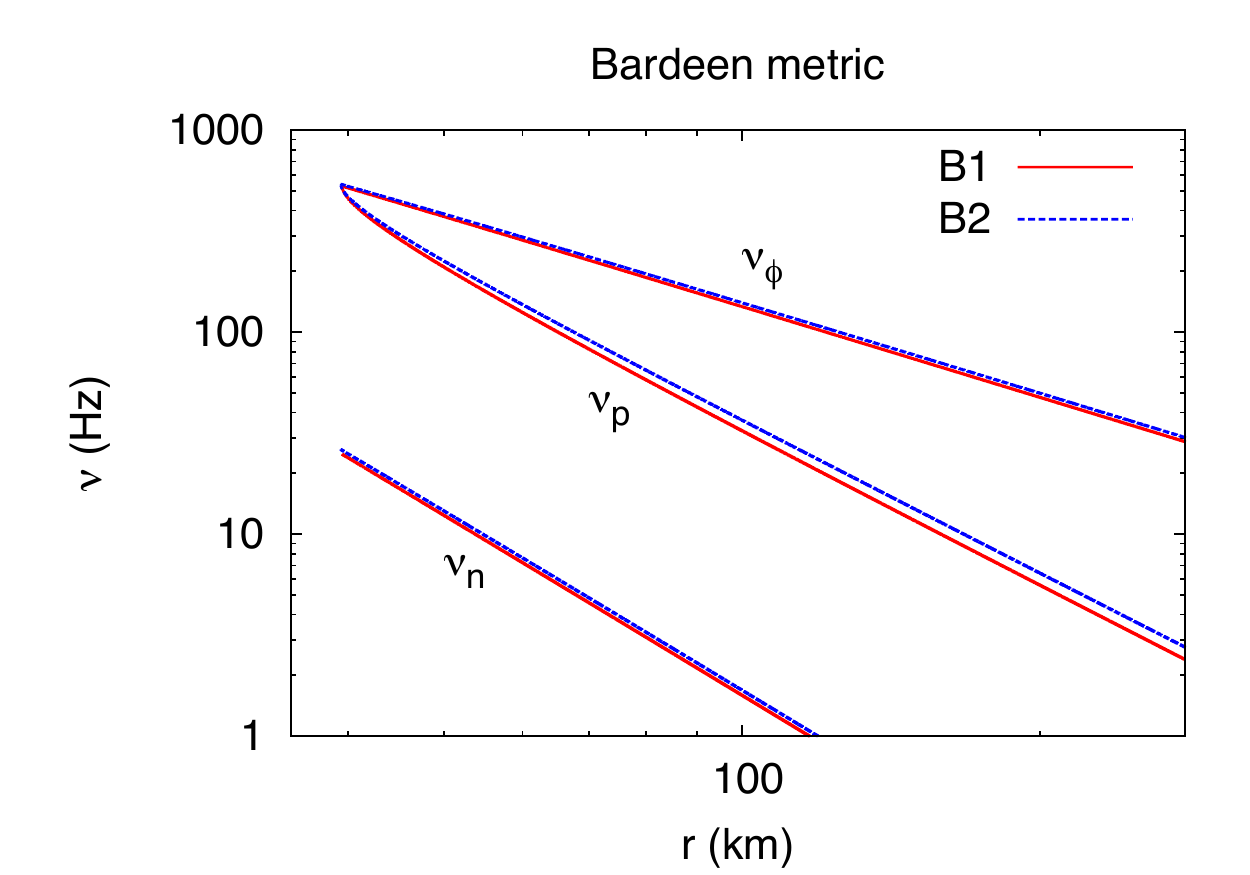}
\includegraphics[type=pdf,ext=.pdf,read=.pdf,width=8.5cm]{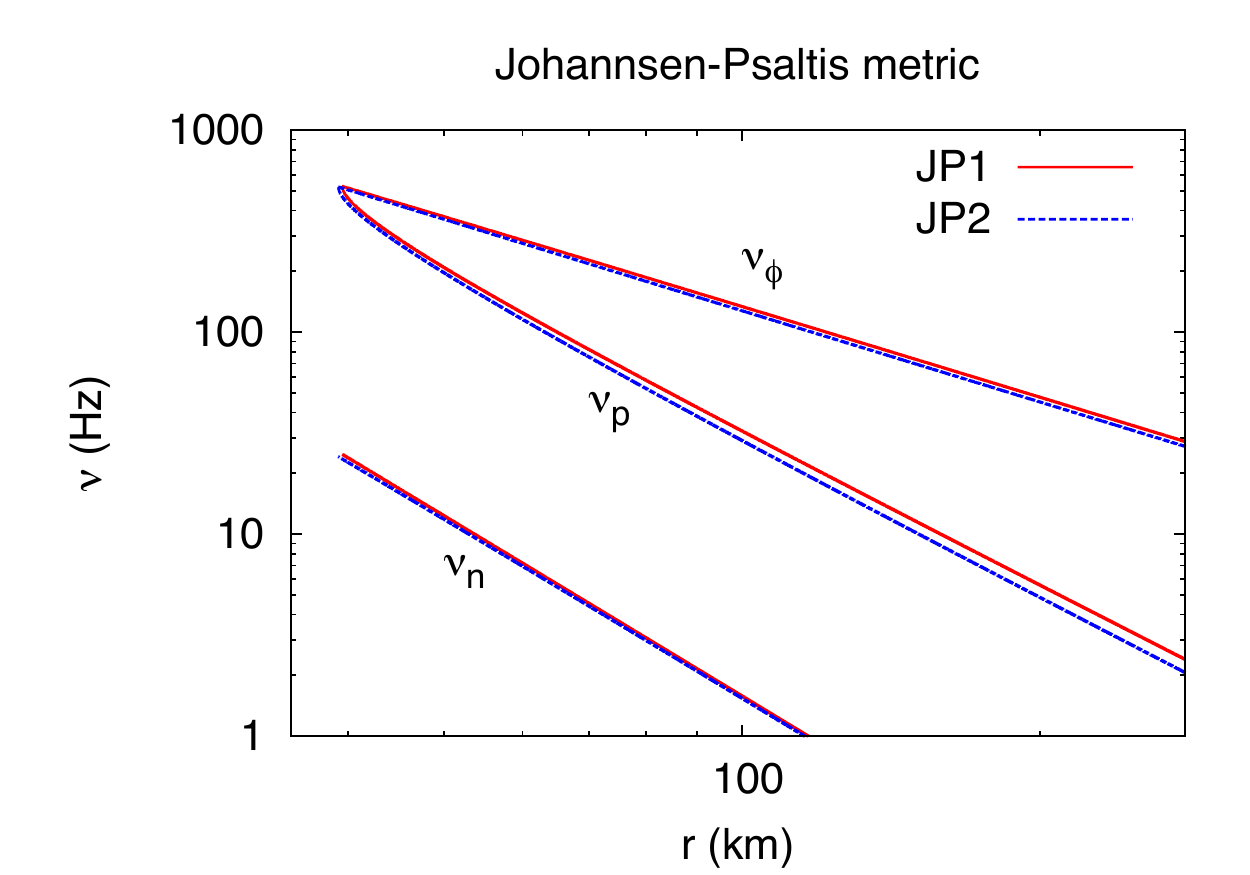}
\end{center}
\vspace{-0.5cm}
\caption{Orbital frequency $\nu_\phi$, periastron precession frequency 
$\nu_{\rm p}$, and nodal precession frequency $\nu_{\rm n}$ as functions
of the orbital radius $r$. Left panel: Bardeen background with $M/M_\odot = 5.40$,
$a/M = 0.279$, and $g/M = 0.23$ (B1) and with $M/M_\odot = 5.95$,
$a/M = 0.243$, and $g/M = 0.56$ (B2). Right panel: Johannsen-Psaltis background 
with $M/M_\odot = 5.42$, $a/M = 0.274$, and $\epsilon_3 = 0.5$ (JP1) and with 
$M/M_\odot = 4.84$, $a/M = 0.339$, and $\epsilon_3 = -2.2$ (JP2). See the text for 
more details.}
\label{fig3}
\end{figure*}

\section{Discussion \label{s-d}}

As shown in Fig.~\ref{fig2}, while the relativistic precession interpretation of 
the data of GRO~J1655-40 is perfectly consistent with the hypothesis that the
spacetime around the BH candidate in this source is described by the Kerr 
metric, large deviations from the Kerr solutions are also allowed. One may 
wonder whether it is possible to solve the tension between the measurement 
inferred from this approach and the one obtained by the continuum-fitting
method in Ref.~\cite{shafee}. The 
estimate of the spin parameter found in the Kerr background in~\cite{shafee} 
can be quickly translated in an allowed region on the spin-deformation 
parameter plane by exploiting the fact that (at least for not too large 
deformation parameters) the disk's thermal spectrum around a deformed 
object with a certain spin is extremely similar to the one of a Kerr BH with 
different spin. Indeed, if we consider a non-Kerr BH metric and we fix the 
value of the deformation parameter, we can find a one-to-one correspondence
between one of these objects and a Kerr BH whose disk's thermal spectrum
is very similar.

With this spirit, if in the Kerr case the allowed spin parameter range is 
$0.6 < a/M < 0.8$, one can just find for any non-vanishing deformation 
parameter the spin of the non-Kerr BH with spectrum similar to a Kerr BH 
with $a/M = 0.6$ and $a/M = 0.8$. The result is the boundary of the
allowed region in the spin-deformation parameter plane, which is
the orange dashed-dotted line in Fig~\ref{fig2}. Here the comparison
of the spectra has been done using the $\chi$-square procedure of,
for instance, the second paper in Ref.~\cite{review}. The fact that there is 
not overlap between the allowed regions suggested by the relativistic 
precession approach and by the continuum-fitting method simply means 
that the tension between the two measurements cannot be solved 
assuming a different spacetime. One arrives
at the same conclusions if the deformation parameter $\epsilon_3$
of the Johansenn-Psaltis background is replaced by higher order
deformations. That has been explicitly checked for $\epsilon_4$, $\epsilon_5$, 
$\epsilon_6$, and $\epsilon_7$, and the general trend suggests it is
correct for any $\epsilon_k$. While it is not possible to firmly exclude 
the possibility that some non-Kerr background can solve the tension 
between the two measurements, the failure of all these attempts suggests 
that such a possibility is at least not very natural.

For GRO~J1655-40, in the literature there are also some estimates of its spin 
parameter with the K$\alpha$ iron line method~\cite{miller}. While the available 
data are not very good, these studies suggest that, in the case of a Kerr BH, the 
object would have a quite high value of the spin parameter, at the level of
$a/M \sim 0.9$ or even higher. It seems thus that the three approaches 
(relativistic precession model, disk's spectrum, iron line) give very different results. 
Following the study of Ref.~\cite{cb3}, it is easy to conclude that for 
the Bardeen metric it is not possible to 
fix the tension between the three measurements, which continue to provide 
three different spins for any value of $g/M$. In the case of the Johannsen-Psaltis 
solution, the results from the continuum-fitting and the iron line analysis may
be consistent in the case of a negative $\epsilon_3$~\cite{cb3}. The compatibility 
between the relativistic precession interpretation and the other 
approaches seems however to be impossible.

Let us now consider what happens if we consider the mass measurement
$M/M_\odot = 6.3 \pm 0.3$ reported in~\cite{shafee},
which is not consistent with the one of~\cite{beer}. If we use this value as input
parameter in the relativistic precession model, we find the plots in Fig.~\ref{fig2bis},
respectively for the Bardeen (left panel) and Johannsen-Psaltis (right panel) 
backgrounds. In the Bardeen metric, the relativistic precession model and the
disk's thermal spectrum are still inconsistent. In the Johannsen-Psaltis spacetime,
we find an overlap between the 2-standard deviation region of the relativistic
precession model and the 1-standard deviation limit of the continuum-fitting 
method. The measurement of the relativistic precession model is
\be
a/M &=& 0.20 \pm 0.04 \, , \nonumber\\
\epsilon_3 &=& 3.8^{+2.4}_{-2.1} \, ,
\ee
at 68\%~C.L. Such a measurement is consistent with the Kerr BH hypothesis
within a 3-standard deviation limit (not within 1- and 2-standard deviation limits), 
but in combination with the analysis of the thermal spectrum of the disk favors a
non-vanishing deformation parameter at the level of $\epsilon_3 \sim 7$.
It is worth noting that the same value was found in the second paper in~\cite{jet} 
from the combination 
of the measurements of the continuum-fitting method and of the power of steady 
jets for 5~BH candidates.

Lastly, it is important to stress that future X-ray satellites like LOFT can have
the capabilities to test the relativistic precession model and hopefully provide 
robust and strong constraints on the nature of stellar-mass BH candidates.
In particular, it would be extremely useful to have observations of QPOs at 
different radii. The three simultaneous QPOs in the available data seem to
occur at a small radial coordinate, $r \sim 45$~km, which corresponds to
a radius close to the innermost stable circular orbit of these spacetimes.
Fig.~\ref{fig3} shows the orbital frequency $\nu_\phi$, the periastron precession 
frequency $\nu_{\rm p}$, and the nodal frequency $\nu_{\rm n}$ as a function 
of the radial coordinate $r$ for the Bardeen and Johannsen-Psaltis backgrounds,
respectively left and right panels. In each panel, the red solid lines are the
fundamental frequencies for the object at the point $\chi^2_{\rm min}$ 
in Fig.~\ref{fig2}. $M/M_\odot = 5.40$, $a/M = 0.279$, and $g/M = 0.23$ 
for the Bardeen case (B1) and $M/M_\odot = 5.42$, $a/M = 0.274$, and 
$\epsilon_3 = 0.5$ for the Johannsen-Psaltis metric (JP1). The blue dashed lines
are instead the frequencies of an object on the 68.3\% C.L. curve in Fig.~\ref{fig2}.
For the Bardeen solution (B2), the parameters are $M/M_\odot = 5.95$, 
$a/M = 0.243$, and $g/M = 0.56$, and they belong to the object with maximum value
of $g/M$ at 68.3\% C.L. For the Johannsen-Psaltis metric (JP2), the parameters are
$M/M_\odot = 4.84$, $a/M = 0.339$, and $\epsilon_3 = -2.2$, and they are associated
to the BH with lowest possible value of $\epsilon_3$ allowed at 68.3\% C.L.
As shown in Fig.~\ref{fig3}, the values of $\nu_\phi$ and $\nu_{\rm n}$ for 
different objects are similar even at larger radii, while the periastron precession 
frequency $\nu_{\rm p}$ seems to be more sensitive to the background metric. 
Very precise measurements of these frequencies at small and large radii may 
thus be an very powerful tool to distinguish Kerr BHs from other BH solutions.

\section{Summary and conclusions \label{s-c}}

Astrophysical BH candidates are thought to be the Kerr BHs predicted in general 
relativity because they are so massive, compact, and dark that they cannot be 
explained otherwise without introducing new physics. Nevertheless, there are not 
yet observations capable of confirming this hypothesis. The properties of the
electromagnetic radiation emitted by the gas in the inner part of the accretion 
disk can potentially provide information on the spacetime geometry around 
these compact objects and thus either confirm the predictions of general relativity
or demand new physics. At present, there are two relatively robust techniques
to probe the metric of BH candidates; that is, the study of the disk's thermal 
spectrum and the analysis of the profile of the K$\alpha$ iron line. However,
these techniques can usually constrain only a certain combination of the spin
parameter and of possible deviations from the Kerr solution, because a non-Kerr
object with a certain spin can likely mimic a Kerr BH with different spin.

In the present paper, I have reconsidered the interpretation of the three QPOs
simultaneously detected in the X-ray data of GRO~J1655-40 proposed in 
Ref.~\cite{belloni} to test the Kerr nature of the stellar-mass BH candidate in
this source. In the Kerr background, the fundamental frequencies associated
to the motion of a test-particle depend only on the orbital radius $r$, the BH
mass $M$, and the spin parameter $a$. Since three QPOs are observed at 
the same time, one can argued that they may be generated at the same
orbital radius and thus solve the system of three equations for the three
fundamental frequencies to find the three variable, $r$, $M$, and $a$. The 
measurement of the mass $M$ found with this approach is consistent with 
the one inferred by studying the orbital motion of the companion star with
optical observations found in~\cite{beer}, but with a smaller uncertainty. 
However, in the literature there is also a different measurement reported 
in~\cite{shafee} and that would be inconsistent with the mass value inferred 
in~\cite{belloni}. The relativistic precession model provides also an
estimate of the BH spin with quite high precision, but it turns out to be in 
disagreement with the value found from the analysis of the disk's thermal 
spectrum and of the iron line profile.

The relativistic 
precession interpretation of the QPOs can potentially be a quite powerful
tool to test the nature of astrophysical BH candidates. In this case, one has
to use the mass $M$ inferred from the optical data  as an independent 
measurement and thus solve the system of three equations for the fundamental
frequencies of the spacetime to find the orbital radius $r$, the spin parameter
$a$, and constrain possible deviations from the Kerr background through the
determination of the deformation parameter under consideration. The data
of GRO~J1655-40 may be consistent with a Kerr BH, but they also 
allow for significant deviations from the Kerr solution. With the mass
measurement of~\cite{beer}, the
disagreement between the results of the relativistic precession interpretation
and the measurement obtained with the continuum-fitting method persists even
relaxing the Kerr BH assumption, and for any choice of the deformation
parameter. 
With the mass measurement of Ref.~\cite{shafee}, the relativistic precession 
model and the continuum-fitting method can be consistent 
in the Johannsen-Psaltis background with non-vanishing $\epsilon_3$
The required deformation is $\epsilon_3 \sim 7$. It is worth noting that the same 
value was found in the second paper in~\cite{jet} by combining the measurements of the 
continuum-fitting method and of the power of steady jets.


\begin{acknowledgments}
This work was supported by the NSFC grant No.~11305038, 
the Shanghai Municipal Education Commission grant for Innovative 
Programs No.~14ZZ001, the Thousand Young Talents Program, 
and Fudan University.
\end{acknowledgments}


\end{document}